\documentclass[prl,aps,twocolumn,groupedaddress,floats,showpacs,final,superscriptaddress]{revtex4}
\usepackage{graphicx}
\usepackage{bm}
\usepackage{color}
\usepackage{epsfig}
\usepackage{psfrag}
\usepackage{amsmath,amsfonts,amssymb,bm,ulem}

\begin{document}

\author{Igor S. Tupitsyn}
\affiliation{Department of Physics, University of Massachusetts, Amherst, MA 01003, USA}
\affiliation{Russian Research Center ``Kurchatov Institute'', 123182 Moscow, Russia}
\author{Andrey S. Mishchenko}
\affiliation{Cross-Correlated Materials Research Group,
RIKEN Advanced Science Institute (ASI), Wako 351-0198, Japan}
\affiliation{Russian Research Center ``Kurchatov Institute'', 123182 Moscow, Russia}
\author{Naoto Nagaosa}
\affiliation{Cross-Correlated Materials Research Group,
RIKEN Advanced Science Institute (ASI), Wako 351-0198, Japan}
\affiliation{Department of Applied Physics, The University of Tokyo,
7-3-1 Hongo, Bunkyo-ku, Tokyo 113, Japan}
\author{Nikolay Prokof'ev}
\affiliation{Department of Physics, University of Massachusetts, Amherst, MA 01003, USA}
\affiliation{Russian Research Center ``Kurchatov Institute'', 123182 Moscow, Russia}
\affiliation{Department of Theoretical Physics, The Royal Institute of Technology, Stockholm SE-10691 Sweden}


\title{Coulomb and electron-phonon interactions in metals}
\date{\today}


\begin{abstract}
An accurate and consistent theory of phonons in metals requires that all long-range Coulomb
interactions between charged particles (electrons and ions) be treated on equal footing.
So far, all attempts to deal with this non-perturbative system were relying on
uncontrolled approximations in the absence of small parameters.
In this work, we develop the Diagrammatic Monte Carlo approach
for a two-component Coulomb system that
obtains the solution to this fundamental problem in an approximation free way by
computing vertex corrections from higher-order skeleton graphs.
The feasibility of the method is demonstrated by calculating
the spectrum of longitudinal acoustic phonons in a simple cubic lattice, determining their sound velocity, and obtaining
the phonon spectral densities by analytic continuation of the Matsubara Green's functions.
Final results are checked against the lowest-order fully self-consistent GW-approximation
in both adiabatic and non-adiabatic regimes.
\end{abstract}


\maketitle

Standard theory of electron-phonon interaction (EPI) in metals involves
a number of approximations. While some of them are based on the small adiabatic
parameter $\gamma=\omega_D/\epsilon_F \sim \sqrt{m/M} \ll 1$ (where $\omega_D$ is the
Debye frequency, $\epsilon_F$ is the Fermi energy, $m$ and $M$ are the electron and
ion masses, respectively), other approximations, such as neglecting (i)
vertex corrections based on the effective electron-electron interaction and (ii) the
mutual self-consistent feedback between the phonon and electron subsystems, remain uncontrolled.
Both effects do not involve small parameters because EPI in metals is inseparable
from strong Coulomb forces between the electrons. Indeed, at the level of the {\it bare} Hamiltonian,
unscreened Coulomb ion-ion interactions prevent formation of longitudinal acoustic phonons
by shifting their frequencies all the way up to the frequency of ionic plasma oscillations,
$\omega_p = q \sqrt{(4\pi e^2/q^2\epsilon_{\infty} ) (n_e /M)}
\equiv \sqrt{ 4\pi n_e e^2/M\epsilon_{\infty}}$,
where $n_e=n_i$ is the conduction electron/ion charge density and
$\epsilon_{\infty}$ is the ion-core dielectric constant. Once
{\it both} the long-range electron-phonon and electron-electron interactions
are accounted for, the acoustic spectrum is recovered back due to screening \cite{Brovman};
the underlying mechanism can be illustrated by replacing $4\pi e^2/q^2$ with
$4\pi e^2/(q^2+\kappa^2)$ (where $\kappa$ is the Thomas-Fermi wavevector)
in the phonon spectrum to get $\omega (q\to 0) \to \omega_p (q/\kappa)$.

In the adiabatic approximation it is assumed that interactions between
(and with) the heavy ions are screened by the static dielectric function
of a metal and the phonon spectrum is determined from the corresponding dynamic matrix 
of a solid. Thus transformed crystal vibrations and EPI 
are no longer singular at small momenta. When further progress is made by separating
effects of electron-electron and electron-phonon interactions \cite{Migdal,Eliashberg},
double-counting is dealt with by excluding static electronic polarization
terms from the renormalization of phonon propagators, and vertex corrections based on
EPI are neglected because they are small in $\gamma$. The adiabatic approximation breaks 
down when $\gamma \gtrsim 1$ is considered, for instance, to explain enhancement of the
critical temperature in phonon-mediated superconductors \cite{Cappelluti,Paci,Pisana}.

However, regardless of the $\gamma$ parameter value, EPI does not involve
natural small parameters in metals and remains strong. This means that even
the first step in the adiabatic approximation (screening of long-range interactions)
is ill-defined since the static dielectric function itself should be the outcome of
the non-perturbative calculation based on all relevant interactions, including EPI.
The importance of vertex corrections was studied by various groups in connection with
superconducting \cite{Grimaldi95,Cappelluti96,Scalapino03,Bauer11} and Dirac 
\cite{DasSarma14} materials, as well as for polarons \cite{Capone03,MiNaPr14},
but never for a two-component Coulomb system in a systematic way when {\it all} forces
are treated on an equal footing, and all uncertainties are quantified.

In this Letter, we develop the bold-line Diagrammatic Monte Carlo (BDMC) technique
that allows us to deal with Coulomb interactions in a fully self-consistent,
approximations free, manner and obtain final results with controlled accuracy by 
accounting for vertex corrections from higher-order skeleton diagrams. We demonstrate 
that BDMC leads to a theory capable of solving the fundamental problem of the phonon 
spectrum in a metal at any $\gamma$, including the most difficult regime of $\gamma 
\sim 1$, i.e. when there are no small parameters of any kind.

Our model simulation considers a simple cubic lattice of vibrating ions coupled to conduction
electrons, and aims at computing the spectrum of longitudinal phonons and their velocity of
sound in the thermodynamic limit. We also perform spectral analysis of the phonon Matsubara 
Green's function in the most difficult parameter regime $\gamma \sim 1$. We show that vertex
corrections to the lowest-order (GW) approximation significantly soften the sound velocity
at $\gamma \gtrsim 1$, and reduce the amplitude of the giant Kohn anomaly at small $\gamma$. 
As far as we know, this kind of studies were not possible to perform in the past;
none of the previous work for Coulomb systems was done at the level of high-order skeleton
technique in the absence of small parameters.

{\it System.} We consider a lattice model of a metal defined by the Hamiltonian
\begin{equation}
H = H_{FH} + H_{c} + H_{ph} + H_{el-ph} \, ,
\label{H_tot}
\end{equation}
where $H_{FH}$ is the standard Fermi-Hubbard model parameterized by
the n.n. hopping amplitude $t$
(with the tight-binding dispersion relation $\epsilon ({\mathbf k})$),
the on-site repulsion $U$, and the chemical potential $\mu$. In what follows
we use the lattice constant $a$ and hopping $t$ as units of length and energy,
respectively.

The second term describes Coulomb electron-electron
interaction, $H_{c} =  \sum_{{\mathbf i} < {\mathbf j}, \sigma, \sigma'} V_c({\bf r}_{ij}) \;
n_{{\mathbf i} \sigma} n_{{\mathbf j} \sigma'}$,
where $n_{{\mathbf i} \sigma}=a^{\dag}_{{\mathbf i} \sigma} a^{\!}_{{\mathbf i} \sigma}$
is the electron density operator for the spin component $\sigma=\uparrow, \downarrow$ on site
${\mathbf i}$ (we employ standard second-quantization notations for creation
and annihilation operators), and $V_c({\bf r}_{ij}) = U_c/|{\bf i}- {\bf j}|$;
in Fourier space, $V_c (q \to 0 ) = 4\pi U_c/q^2$.
We consider $U_c=e^2/\epsilon_{\infty}$ as an independent (from $U$) parameter;
the bare electron-electron interaction is defined as the sum of local (spin-dependent)
and non-local terms:
$V_{ee}({\bf r}_{ij}) = U \delta_{{\bf r}_{ij}} + V_c({\bf r}_{ij})$
(for brevity, we do not explicitly mention the tensor structure of interactions,
propagators, and irreducible objects in the spin space).

The Hamiltonian of ionic system is assumed to be harmonic and described by
a collection of longitudinal phonons \cite{transverse},
$H_{ph}=\sum_q \omega({\bf q}) b^{\dag}_{\bf q} b_{\bf q}$. Their
bare spectrum is gapped at small momenta: $\omega ({\mathbf q}\to 0)=\omega_p$.

The electron-ion interaction has the standard density-displacement form,
\begin{equation}
H_{el-ph} = i\sum_{{\bf q}, {\bf k}, {\sigma}} M({\bf q}) \;
a^{\dag}_{{\bf q+k},\sigma} a^{\!}_{{\bf q}, \sigma}
(b^{\!}_{\bf k} + b^{\dag}_{\bf -k}) \, ,
\label{el-ph}
\end{equation}
with the interaction vertex $M({\bf q})$  based on the derivative of the Coulomb
electron-ion potential. Since in all expressions we {\it always} have to deal with
$|M({\mathbf q})|^2$, it makes sense to introduce $V_{ep}({\mathbf q})=|M({\mathbf q})|^2$
whose asymptotic form  $[\omega_p/2] V_c (q)$ at $q \to 0$ is unambiguously fixed
by electro-neutrality of the system.
Solely for the purpose of minimizing the number of model parameters, we confine ourselves to
$\omega ({\mathbf q}) = \omega_p$ and $V_{ep}(r)=(\omega_p/2) V_{ee}(r)$
with $\omega_p/t=0.5$.

\begin{figure}[h]
\vspace{-3.2cm}
\includegraphics[width=7cm]{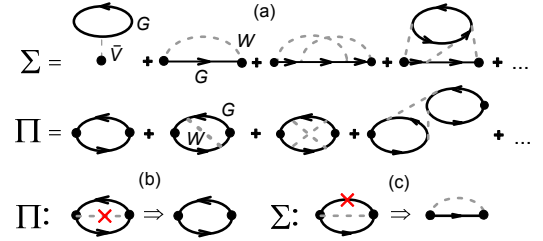}
\vspace{-3.2cm}
\caption{ (a) Skeleton (irreducible) diagrams for electron self-energy $\Sigma$ and
polarization function $\Pi$ in terms of fully dressed Green's functions $G$ and screened
interactions $W$. Hartree term is the only graph based on the bare potential $\bar{V}$.
(b-c) To go from free-energy diagrams to those for $\Pi$ (b) or $\Sigma$ (c),
one has to remove the ``measuring" line marked by the red cross.}
\label{Fig1}
\end{figure}

{\it Methodology}. Our calculations are based on the so-called $G^2W$-expansion, see
Fig.\ref{Fig1}(a), when irreducible (with respect to cutting one line) diagrams for
self-energy $ \Sigma $ and polarization $\Pi$ are expressed in terms of
fully-dressed Green's functions, $G$, and screened  effective
interactions, $W$, defined
self-consistently through Dyson equations in the Matsubara frequency-momentum space:
\begin{equation}
G^{-1} = G_0^{-1}-\Sigma \;, \qquad \qquad  W^{-1} = \bar{V}^{-1}-\Pi \;.
\label{Dyson}
\end{equation}
Here $G_0^{-1} = i\omega_n + \mu -  \epsilon ({\mathbf k})$ and
$D_0^{-1} = [ \omega_m^2 + \omega^2({\mathbf q}) ] / 2\omega ({\mathbf q})$,
are the bare electron and phonon Green's functions, respectively.
Their Matsubara frequencies are defined differently:
for fermions,  $\omega_n = 2\pi T (n+1/2)$ with integer $n$; for bosons,
$\omega_m = 2\pi T m$ with integer $m$. Within the $G^2W$-expansion framework,
one has to combine the bare electron-electron potential with the phonon-mediated
term to form the frequency-dependent potential $\bar{V} = V_{ee}-D_0V_{ep}$ appearing in
the second Dyson equation. This formulation is complete in a sense that
exponential convergence of the skeleton sequences with increasing the diagram
order leads to the final solution of the problem \cite{Riccardo}.

To determine properties of the phonon subsystem, we define the polarization
function irreducible with respect to cutting one phonon line,
$\Pi^{-1}_{P} = \Pi^{-1}-V_{ee}$. By construction, in combination with
$V_{ep}$, it plays the role of self-energy for the renormalized phonon propagator
\begin{equation}
D^{-1} = D_0^{-1}-\Sigma_{ph}; \qquad \qquad \Sigma_{ph}=-V_{ep}\Pi_{P}  \;.
\label{DysonP}
\end{equation}

Our implementation of the BDMC technique is closely following that described in
Ref.~\cite{KULAGPRL2013}. We sample the configuration space of
\cite{irreducible} skeleton free-energy diagrams in the $({\bf r}, \tau)$-representation
with one of the lines always being marked
(by red cross in Fig.\ref{Fig1}(b-c)) as ``measuring";
its functional dependence on space-time coordinates of its end-points is arbitrary.
When the ``measuring line" is removed, the remaining diagram contributes
either to $\Pi$, see Fig.~\ref{Fig1}(b), or to $\Sigma$, see Fig.\ref{Fig1}(c).
In the imaginary-time representation, we need to split $W$ into the sum of the
bare electron-electron potential,
$V_{ee}({\bf r}_{ij}) \delta (\tau_1-\tau_2)$, and the rest,
$W-V_{ee}({\bf r}_{ij}) \delta (\tau_1-\tau_2)$,
because $\delta$-functional and generic functional dependencies on time
are incompatible. This implies, in particular, that the measuring
line cannot be of the $V_{ee}$-type \cite{KULAGPRL2013}.

Both $\Sigma$ and $\Pi$ are computed as sums of skeleton graphs, up to order $N$
(there are $2N$ vertexes in the $N$-th order graph); we will denote these sums as
$\Sigma_N$ and $\Pi_N$.
The lowest-order contributions are known right away because they are nothing but products of
$G$ and $W$ functions; in the skeleton formulation, $\Sigma_1$ and $\Pi_1$ are equivalent to the
GW-approximation with fully self-consistent treatment of the EPI feedback on polarization.
By charge neutrality, Hartree terms involving $V_{c} ({\bf r}\ne 0)$ have to be removed.
Thus, Monte Carlo statistics has to be collected only from
higher-order diagrams and then added to the GW-result. The self-consistency loop is
closed after $\Sigma$ and $\Pi$ are used in Dyson equations to define new $G$ and $W$
functions that are subsequently considered in all diagrams as the simulation continues.
To solve Dyson equations (\ref{Dyson}-\ref{DysonP}) we employ Fast-Fourier-Transform
algorithms to go to the momentum-frequency space where these equations
are algebraic.

\begin{figure}[h]
\vspace{-3.7cm}
\includegraphics[width=8.5cm]{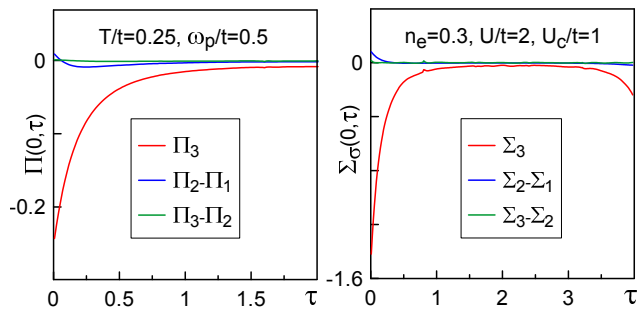}
\vspace{-3.7cm}
\caption{Convergence properties of the skeleton sequence for system size $L=16$
(other parameters are specified in the legends). In the left panel the local
polarization $\Pi(0,\tau) = \sum_{\sigma}\Pi_{\sigma \sigma}(r=0,\tau)$ and its
order-by-order contributions are shown as functions of $\tau$ to demonstrate that
$\Pi_3$ result (red line) is an order of magnitude larger than its partial contribution
from diagrams of the second order (blue); third-order diagrams make an even smaller
contribution (green line). In the right panel an identical analysis is presented for
local self-energy $\Sigma_{\sigma}(0,\tau) = \Sigma_{\uparrow}(0,\tau)
= \Sigma_{\downarrow}(0,\tau)$, with the same color scheme meaning. }
\label{Fig2}
\end{figure}

The largest system size simulated in this work was $L^3=64^3$, with periodic
boundary conditions. The thermodynamic limit was recovered by extrapolating
results obtained for $L=16$, $32$, and $64$ to infinity. We also have to perform
an extrapolation to the $N\to \infty$ limit, or observe good
convergence of results with increasing $N$. In two panels of Fig.\ref{Fig2} we plot
local polarization $\Pi(r=0,\tau)$ and self-energy $\Sigma_{\sigma}(r=0,\tau)$
along with their partial order-by-order contributions. Clearly, contributions from the
third-order skeleton graphs are already very small, but understanding their
role is required for estimating accuracy limits of calculations truncated at $N=2$.

{\it Results.} The tight-binding model on a simple cubic lattice at half-filling satisfies 
the ``nesting" condition at momentum ${\bf Q}_N=(\pi,\pi,\pi)$. This leads to singularity
in the density of states, logarithmic divergence of polarization at zero temperature,
$\Pi({\bf q}\to {\bf Q}_N) \sim \ln|{\bf q}-{\bf Q}_N|$, and the corresponding ``giant"
Kohn anomaly in the phonon spectrum (typical for one-dimensional systems) \cite{Kohn1959}.
It is expected then that at low temperature the phonon spectrum is anomalously
soft at ${\bf Q}_N$ and there is a structural phase transition with the dominant
density modulation at ${\bf Q}_N$. In contrast, the conventional Kohn anomaly is linked 
to the logarithmic divergence of the polarization derivative $\partial \Pi / \partial q$ 
at momentum transfer ${\bf q}=2{\bf k}_F$  (at $T=0$).

\begin{figure}[h]
\vspace{-1.2cm}
\includegraphics[width=6cm]{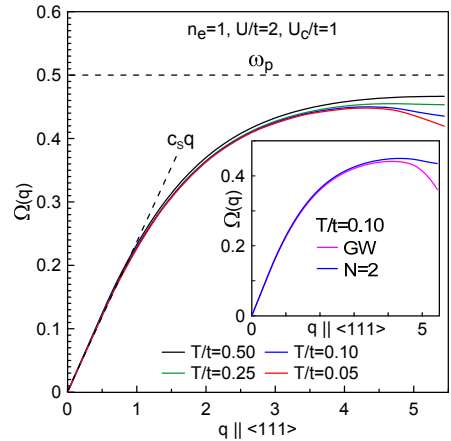}
\vspace{-1.2cm}
\caption{Giant Kohn anomaly in the renormalized phonon spectrum $\Omega({\bf q})$ at the
nesting vector ${\bf Q}_N=(\pi ,\pi ,\pi )$ for $L=32$, $n_e=1$, and $N=2$.
In the inset we show how $\Omega({\bf q})$ at $T/t=0.1$ depends on the diagram order:
GW-approximation (magenta), $N=2$ (blue).}
\label{Fig3}
\end{figure}

In Fig.~\ref{Fig3} we show the dramatic temperature dependence of the renormalized phonon
dispersion $\Omega({\bf q})$ (along the $\langle111\rangle$ direction) at half-filling.
The spectrum was deduced from the pole-approximation $D^{-1} \propto \omega^2-\Omega^2({\bf q})$
for the phonon propagator, see Eqs.(\ref{DysonP}). As temperature decreases, the cusp at
${\bf Q}_N$ is getting more pronounced and the phonon spectrum softens;
temperature scales (and appropriate system sizes) required for studying
the structural transition point are exponentially small (large) in this case.
Vertex corrections substantially reduce the amplitude of the Giant Kohn anomaly,
see inset in Fig.~\ref{Fig3}, but do not eliminating it.
Within the GW-approximation the phonon spectrum goes unstable at $T \lesssim 0.03$.

\begin{figure}[h]
\vspace{-1.2cm}
\includegraphics[width=6cm]{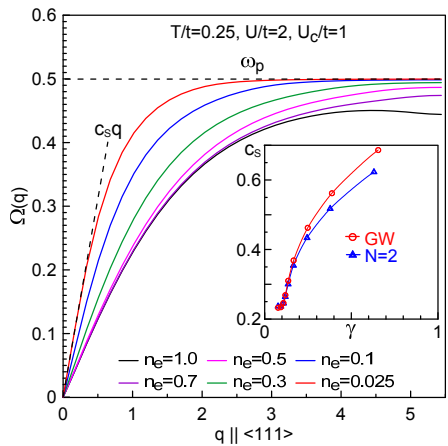}
\vspace{-1.2cm}
\caption{ Renormalized phonon dispersion $\Omega({\bf q})$ along the $\langle111\rangle$
direction for various electron densities at $T=\omega_p/2 = 0.25t$.
All results were obtained for $L^3=64^3$ and $N=2$. Inset: sound velocity
as a function of $\gamma$ within the GW-approximation (red) and with $N=2$ vertex
corrections (blue). Both curves are extrapolated to the thermodynamic limit
from the $L=16, \; 32, \; 64$ set. Error bars are smaller than symbol sizes. }
\label{Fig4}
\end{figure}

Away from half-filling, the phonon spectrum should demonstrate the standard Kohn anomaly
at $q=2 k_F$ smeared by finite-temperature effects. It can be seen as a small wiggle
on the phonon dispersion curve corresponding to density $n_e=0.7$ in the momentum
interval $4.5 < q < 5$ (at this filling factor, $k_F \approx 2.4$ along the $\langle111\rangle$
direction), see the main plot in Fig.\ref{Fig4}.

As far as screening effects are concerned, the plasmon gap at $q\to 0$ is closed at
all densities, and $\Omega({\bf q}\to 0)$ clearly demonstrates the characteristic
sound-wave dependence $c_s q$, see Figs. \ref{Fig3} and \ref{Fig4}.
When $n_e$ decreases (at constant $U_c$ and $\omega_p$ this implies that ions are getting
lighter, $M = (4\pi U_c /\omega_p^2) n_e$) the spectrum at large values of $q>2k_F$ saturates
at $\omega_p$, see the main panel in Fig.\ref{Fig4}, and the sound velocity increases, see
Fig.\ref{Fig4} inset. This behavior is in complete agreement with the Fermi-liquid theory
prediction $c_s \propto (k_F/m) \sqrt{m/M} \propto n_e^{-1/6}$ at constant plasma frequency.

Near half-filling, $0.5 < n_e < 1$, where the adiabatic parameter is small,
$\gamma \lesssim 0.1$, the effect of higher-order vertex corrections on sound
velocity appears to be small, and phonon spectra at small momenta are
indistinguishable within the error bars (this is not the case for large momenta,
especially at ${\bf Q}_N$, see the inset in Fig.\ref{Fig3}). As expected,
higher-order diagrams start playing a role at low density when the adiabatic parameter
is approaching unity. In the inset of Fig.\ref{Fig4} we show how the sound velocity
depends on $\gamma$ and how strong the effect of vertex corrections is.
At densities $n_e \lesssim 0.2$  (or $\gamma \gtrsim 0.2$) the GW-approximation
becomes rather unsatisfactory.

\begin{figure}[h]
\vspace{-2.1cm}
\includegraphics[width=8cm]{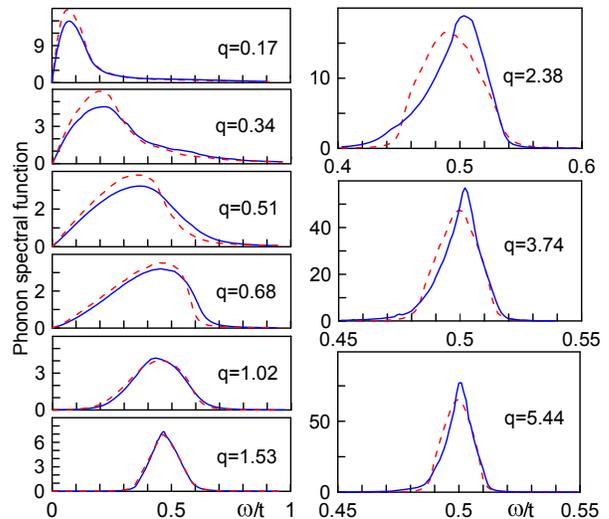}
\vspace{-2.1cm}
\caption{ Phonon spectral functions within the GW (blue solid lines) and $N=2$
(red dashed lines) approximations at $T=0.25t$ and $n_{e}=0.025$ for $L^3=64^3$.}
\label{Fig5}
\end{figure}

To gain additional information on dynamic properties of phonons,
we perform analytic continuation of the Matsubara Green's function
$D({\bf q},\tau)$ (with $\Pi (q=0, \omega_m ) \propto \delta_{m,0}$ obeying
the particle conservation law requirement) into the real frequency domain.
This is done by a combination of the unbiased stochastic optimization and consistent
constraints methods \cite{Andrey2000,PS13}. In Fig.\ref{Fig5} we show
the phonon spectral function for several values of momenta
along the $\langle111\rangle$ direction at $T=\omega_p/2 = 0.25t$, and
compare GW with $N=2$ results. This is done in the most difficult low-density
limit $n_e=0.025$ where $\gamma \sim 1$. Note the large width of phonon peaks
that is often comparable to their energies.
Strong damping of longitudinal phonons is an inevitable property accompanying
screening of long range interactions, which has been observed in metals since
early neutron scattering experiments \cite{Brock}.
First, the phonon damping is increasing with $q$ at small momenta, but then
the phonon lines are getting more narrow at larger values of $q$ as the phonon
life-time is becoming longer. For the three largest values of $q > 2k_F$,
the phonon energy (first moment of the spectral function)
saturates at $\omega_p$, in accordance with Fig.\ref{Fig4}.

{\it Conclusions.} We developed and applied the BDMC approach to solve for electronic and
and vibrational properties of a metal in a fully  self-consistent approximations free way
by dealing with all Coulomb interactions on equal footing in the absence of small parameters.
We find that the skeleton sequence converges fast for our parameters,
and if final results are desired with accuracy of the order of one percent then it is
sufficient to account only for the lowest-order vertex corrections in most
cases. To arrive at this conclusion, we had to quantify the contributions from
higher-order graphs. The presented field-theoretical framework allows one to address virtually
any question about system's statistical behavior.

We demonstrated that our calculations capture the essence of screening effects in metals,
and allow precise calculations of the renormalized phonon spectrum and sound
velocity for all values of $\gamma$. In this work we focused on basic principles
and discussed only the longitudinal acoustic phonons; including other phonon branches
is left for future work but we do not see any difficulty in this regard.
One may also quantify the feedback of the phonon subsystem on electronic
properties (spectrum, dielectric function, optical conductivity,
effective interactions, etc.) and aim at computing the irreducible
Cooper-channel couplings. It would be equally interesting to investigate
the relative effect of the on-site repulsion $U$ on all quantities.

{\it Acknowledgements.} We thank B. Svistunov for discussions. This work was supported by
the Simons Collaboration on the Many Electron Problem, the National Science Foundation under
the grant PHY-1314735, the MURI Program ``New Quantum Phases of Matter" from AFOSR, the
Stiftelsen Olle Engkvist Byggm\"{a}stare Foundation, and the Swedish Research Council grant 642-2013-7837.
N.N. and A.S.M. are supported by Grant-in-Aids for Scientific Research (S) (No. 24224009)
from the Ministry of Education, Culture, Sports, Science and Technology(MEXT) of Japan,
and by ImPACT Program of Council for Science, Technology and Innovation
(Cabinet office, Government of Japan).



\begin{thebibliography}{99}
\bibitem{transverse}
We do not consider transverse phonons in this work because they (i)
do not contribute to the discussion of the fundamental role played
by screening effects, (ii) introduce additional model parameters
irrelevant for purposes of this study, and (iii) are trivial to
include in the proposed numerical scheme.
\bibitem{irreducible}
To have in place an efficient (diagram order independent) mechanism for
verifying that the graph is irreducible, each line is assigned an auxiliary
``momentum" and the momentum conservation law is enforced at each vertex.
In irreducible graphs no two lines carry the same momentum---this can be
quickly established with the help of the hash-table trick.
\end{thebibliography}
\end{document}